\documentclass[twocolumn]{aastex631}

\usepackage{enumerate}
\usepackage{color}
\usepackage{soul}
\usepackage{amsmath}

\newcommand{\Msol}{M_\odot}

\accepted{October 13, 2023}
\submitjournal{ApJ}

\shorttitle{Mass ratio estimates using the derivatives of light curves}
\shortauthors{Kouzuma}

\graphicspath{{./}{./}}
\begin{document}

\title{Mass ratio estimates for overcontact binaries using the derivatives of light curves}

\author[0000-0002-0008-9979]{Shinjirou Kouzuma}
\affiliation{Faculty of Liberal Arts and Sciences, \\
Chukyo University, \\
101-2 Yagoto-honmachi, Showa-ku, Nagoya, Aichi 466-8666, Japan}
\correspondingauthor{Shinjirou Kouzuma}
\email{skouzuma@lets.chukyo-u.ac.jp}

\begin{abstract}
The photometric mass ratios of eclipsing binaries are usually estimated by light-curve modeling with an iterative method. 
We propose a new method for estimating the photometric mass ratio of an overcontact binary using the derivatives of a light curve, which provides a reasonable uncertainty value. 
The method mainly requires only the time interval value between two local extrema found in the third derivative of a light curve, with no need of an iterative procedure. 
We applied the method to a sample of real overcontact binary data and compared the estimated mass ratios with their spectroscopically determined values. 
The comparison showed that our estimated mass ratios for $\sim $67 \% of the samples agreed with their spectroscopic mass ratios within the estimated uncertainties, and the errors for 95 \% of them are within $\pm 0.1$. 
Our method should be useful for estimating mass ratios for numerous overcontact eclipsing binaries found with existing and future surveys, as well as for the light-curve analysis of each system. 
\end{abstract}


\section{Introduction} \label{sec:intro}
The mass ratio of component stars is a fundamental parameter in binary systems. 
In an overcontact binary, the two component stars overfill their Roche lobes; the mass ratio is closely associated with the physical properties of the lobes. 
Accordingly, the shape of the binary system's combined surface strongly depends on the mass ratio in addition to the degree of overcontact. 
Furthermore, the absolute parameters of a binary cannot be determined unless the mass ratio is known. 
The mass ratios of overcontact binaries are also known to have close relationships with other binary parameters, such as the ratios of radii and luminosities \citep{Osaki1965-PASJ, Lucy1968-ApJ877}. 
Moreover, the distribution of the mass ratios in populations of binaries is also important for understanding the formation and evolution of binary systems \citep{Trimble1990-MNRAS, Mazeh1992-ApJ, Rucinski2001-AJ1007}. 

The most widely used and reliable method to determine a mass ratio is based on radial velocity observations through spectroscopy. 
The determined mass ratio in this way is called the spectroscopic mass ratio. 
A major downside of the method is that the observations are highly resource-intensive, usually requiring relatively advanced instruments and repeated spectroscopic observations. 
Such a situation hinders obtaining spectroscopic mass ratios for numerous eclipsing binaries \citep{Terrell2022-Galax}. 

A more convenient method to estimate a mass ratio in a binary system is by the use of photometry-based light curves (LCs). 
The ratio obtained through this method is called the photometric mass ratio, and it is described, e.g., in works such as \citet{Kopal1959-cbs} and \citet{Wilson1994-PASP}. 
A popular method to estimate a photometric mass ratio is the so-called $q$-search method; 
a mass ratio ($q$) is determined through the model fitting of observed photometry LCs with a set of trial values for the mass ratio. 
Many authors have estimated such mass ratios for eclipsing binaries that lack radial velocity data \citep[e.g., see the catalog by][and references therein]{Latkovic2021-ApJS}. 
For overcontact systems with total annular eclipses, it has been demonstrated that photometric mass ratios are accurately estimated \citep{Wilson1994-PASP, Pribulla2003-CoSka, Terrell2005-ApSS}. 
However, the derivation of photometric mass ratios requires an appropriate parameter setting and LC modeling with an iterative method. 
Such situations make it difficult to estimate photometric mass ratios for large numbers of eclipsing binaries. 

In this study, we find that a time interval in the third derivative of an LC is closely correlated with the mass ratio. 
Using the relation, we propose a new method to estimate photometric mass ratios, which is straightforward, with no iterative procedures involved, and provides reasonable uncertainty values. 
We apply this method to real binary data and discuss its effectiveness and uncertainties.

\begin{figure*}
\centering
\includegraphics[width=0.98\textwidth]{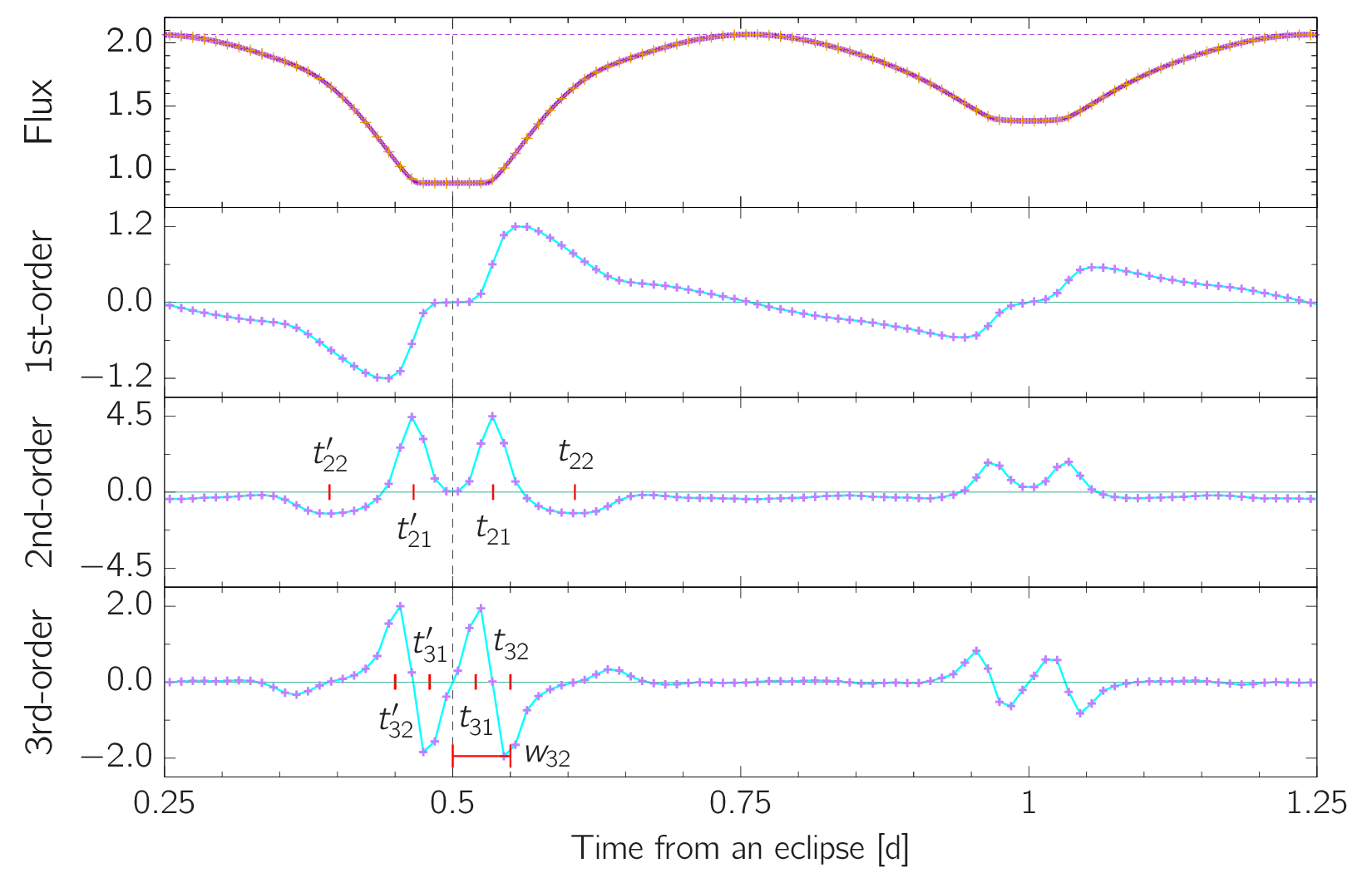}
\caption{Example of a synthetic LC (the uppermost panel) and its first to third derivatives (the second to fourth panels, respectively) for an overcontact binary with $q=0.35$, $P=1$ day, $M_\mathrm{p}=1.5$ $\Msol$, $i=90^\circ$, $T_\mathrm{p}=5000$ K, $T_\mathrm{s}=6000$ K, and $f=0.2$. 
The values of the flux and the first to third derivatives are in units of W m$^{-2}$, $10$ W m$^{-2}$ day$^{-1}$, $10^2$ W m$^{-2}$ day$^{-2}$, and $10^4$ W m$^{-2}$ day$^{-3}$, respectively. 
\label{fig:LC-diff}}
\end{figure*}
\section{Data and Method} \label{sec:method}
\subsection{Synthetic LC samples}
We synthesized a total of 117,600 LCs for overcontact binaries using the software package PHOEBE, release 2.4 \citep{Conroy2020-ApJS}. 
Here, we adopted the following parameter ranges and steps: 
the mass ratio $q=M_\mathrm{s}/M_\mathrm{p}=0.05$--$0.95$ (in steps of 0.1), orbital period $P=0.2$--$1.8$ (0.4) days, primary mass $M_\mathrm{p}=0.5$--$2.0$ (0.5) $\Msol$, orbital inclination $i=60^\circ$--$90^\circ$ ($10^\circ$) deg, primary star and secondary star temperatures $T_\mathrm{p}$ ($T_\mathrm{s}$)$=4000$--$10,000$ (1000) K, and fillout factor $f=0.2$--$0.8$ (0.3). 
The gravity-darkening coefficient was set either to 0.32 or 1, depending on whether the star's temperature was lower or higher than 6600 K, respectively. 
The fluxes were computed at every 0.001 step in phase. 

\subsection{Key values for estimating a mass ratio}\label{sec:key-value}
In our method, we first find the numerical derivatives of an LC up to order 3, with respect to time. 
This work utilizes the second-order central difference for the mean fluxes, each data point of which is 100 phase bins combined. 
We find a higher-order derivative for each by repeating numerical differentiation. 

Figure \ref{fig:LC-diff} shows an example of an LC and its numerical derivatives with respect to time. 
Here, we define the symbol $t_{ij}$ as the time at either a local maximum or minimum. 
This method requires the following conditions: 
\begin{enumerate}
	\item The second derivative of the LC has two local maxima at $t_{21}$ and $t'_{21}$ that are symmetric with respect to an eclipse time. 
	\item The third derivative of the LC has a local  maximum at $t'_{32}$ and a local minimum at $t_{32}$. 
Both extrema appear around the eclipse time determined by condition (1) and $w'_{32} \sim w_{32}$, where $w_{ij}$ is the time interval between $t_{ij}$ and the eclipse time. 
	\item The condition $w_{31}<w_{21}<w_{32}<w_{22}<0.2 P$ must be satisfied, where $P$ is the orbital period. 
\end{enumerate}
In some cases, the local extrema at $t_{31}$ and $t'_{31}$ can be difficult to find. 
However, as long as a symmetric double peak as described in condition (1) is found, a failure to identify the local extrema does not affect the result. 

In this study, a key value is the time interval $t_{32}-t'_{32}$. 
We obtained these two times by finding the zero derivatives using linear interpolation between the derivatives at consecutive phase points.  
If the shapes of the two peaks at $t_{21}$ and $t'_{21}$ are unclear in an LC, the derived mass ratio will be accordingly erroneous. 

Next, we compute the following value $W$: 
\begin{eqnarray}\label{eq:W}
	W = \frac{P}{t_{32}-t'_{32}} = \frac{P}{w_{32}+w'_{32}}. 
\end{eqnarray}
Here, when using the phase instead of the orbital period, we substitute unity for $P$. 
If a symmetric double peak (i.e., two local maxima at $t_{21}$ and $t'_{21}$) is found around both eclipses in the second derivative of an LC as in Figure \ref{fig:LC-diff}, 
we use the mean of both the $W$ values and also calculate the uncertainty described in Section \ref{sec:uncertainty} from both the $W$ values. 
If an LC has an inadequately small number of data points or insufficient photometric precision, such a double peak may appear even when it does not actually exist or vice versa. 
As shown in the following sections, this value is closely correlated with the mass ratio and is a key in our method.

\begin{figure}
\centering
\includegraphics[width=0.48\textwidth]{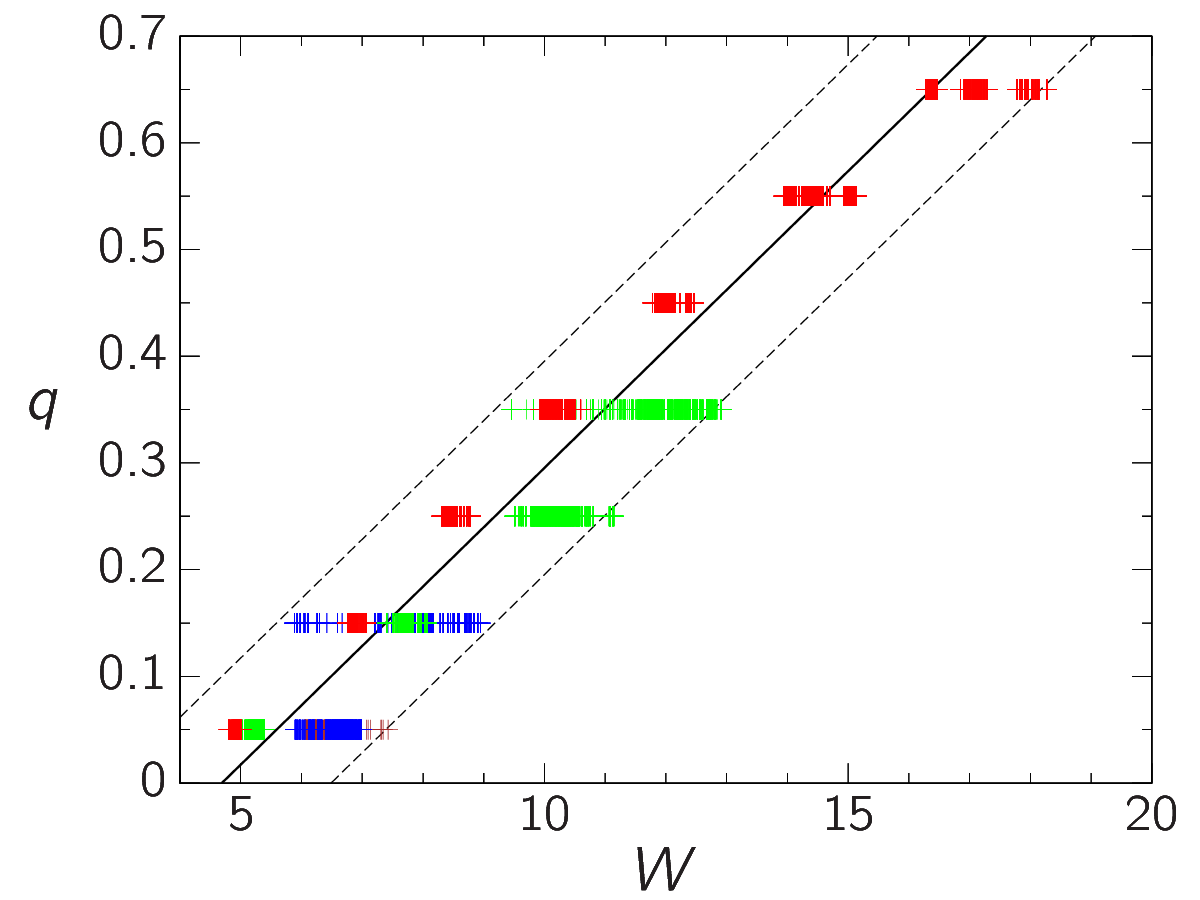}
\caption{The relation between $W$ (Equation (\ref{eq:W})) and mass ratio $q$, calculated from our synthesized LCs (see Section \ref{sec:method}). 
The red, green, blue, and brown crosses represent binaries with $i=90^\circ$, $80^\circ$, $70^\circ$, and $60^\circ$, respectively. 
The solid line is the regression line, accompanied by a pair of parallel dashed lines for reference, with mass ratios of larger and smaller by 0.1 than the regression line. 
\label{fig:q-reg}}
\end{figure}
\subsection{Mass ratio estimation} \label{sec:q-estimation}
In this section, we derive a generic formula for estimating mass ratios. 
We calculated the $W$ values (Equation \ref{eq:W}) for 36,111 binaries out of the 117,600 synthesized LCs. 
Figure \ref{fig:q-reg} shows the relation between the mass ratio $q$ and calculated $W$ values. 
The $W$ values of binaries with inclinations of $i=60^\circ$, $70^\circ$, and $80^\circ$ in our synthesized samples were found to be defined for those with mass ratios of $q=0.05$, $q\leq 0.15$, and $q\leq 0.35$, respectively. 
We found a clear positive correlation between $W$ and $q$. 
A regression analysis for the correlation yielded 
\begin{equation}\label{eq:q-reg}
	q = 0.056 W - 0.261. 
\end{equation}
The $W$ values for 99 \% of our synthesized LCs were found to fall within a mass ratio range of $q \pm 0.1$ (the two dashed lines in Figure \ref{fig:q-reg}). 
This indicates that our method can predict a mass ratio with an error of less than 0.1 typically and $\lesssim$ 0.12 even in the worst case. 

In Figure \ref{fig:q-reg}, a dispersion in the $W$ values is observed for a mass ratio. 
The primary factor for the dispersion is the difference in inclination, although other factors contribute to a lesser degree. 
Also, the amount of dispersion is observed to be an increasing function of the mass ratio, though its contribution to the dispersion is relatively small. 
Note that the fillout factor has a much smaller effect on the dispersion. 
The standard deviation ($\sigma$) of the residuals for $W$ is $\sigma_W = $ 0.766, which corresponds to $\sigma_q = $ 0.043. 
This dispersion could not be reduced, and it contributes to the uncertainty of the estimated mass ratio in this method. 

In addition, Table \ref{tab:std} presents the residual standard deviations of the estimated mass ratios using our method for the synthesized LCs with the mass ratios shown in the table. 
These values offer insights into the extent of the uncertainty in our estimates for the mass ratios of the LCs.

\begin{deluxetable}{CCC}
\tablenum{1}
\tablecaption{Residual standard deviations of the estimated mass ratios\label{tab:std}}
\tablewidth{0pt}
\tablecolumns{4}
\tablehead{
\colhead{\quad \quad \quad \quad$q$\quad \quad \quad \quad} & \colhead{\quad \quad \quad \quad$\sigma_q^-$\quad \quad \quad \quad} & \colhead{ \quad\quad \quad \quad$\sigma_q^+$\quad \quad \quad \quad}
}
\decimalcolnumbers
\startdata
0.05 & -0.029 & 0.058 \\
0.15 & -0.030 & 0.027 \\
0.25 & -0.042 & 0.057 \\
0.35 & -0.048 & 0.057 \\
0.45 & -0.044 & \mathrm{--}  \\
0.55 & -0.023 & 0.023 \\
0.65 & -0.002 & 0.053 \\
\enddata
\tablecomments{
The first column ($q$) displays the mass ratios of synthesized LCs. 
The values of $\sigma_q^-$ and $\sigma_q^+$ were computed only for the binaries for which mass ratios were underestimated and overestimated with the proposed method, respectively. 
The $\sigma_q^+$ value for $q=0.45$ could not be derived because no $W$ values exceeding the regression line were found for the synthesized LCs with $q=0.45$ (see Figure \ref{fig:q-reg}). 
}
\end{deluxetable}

\subsection{Uncertainty}\label{sec:uncertainty}
Equation (\ref{eq:W}) defines that $W$ is a function of $P$, $t_{32}$ ($w_{32}$), and $t'_{32}$ ($w'_{32}$). 
Under the assumption that the uncertainty of $P$ is sufficiently small and that the two local extrema in the third derivative of an LC are perfectly symmetric (i.e., $w_{32}=w'_{32}$), the standard uncertainty of $W$ is 
\begin{eqnarray}
	\delta W = \frac{|w_{32}-w'_{32}|}{P} W^2. 
\end{eqnarray}
Considering the uncertainty of $W$ mentioned in Section \ref{sec:q-estimation} together with $\delta W$, the uncertainty of a mass ratio derived in this method is estimated to be 
\begin{eqnarray}
	\delta q &=& 0.056 \sqrt{\sigma_\mathrm{W}^2 + \delta W^2} \\
	&\simeq & \sqrt{0.043^2 + \left(0.056 \delta W \right)^2}. \label{eq:uncertainty}
\end{eqnarray}
As shown in Section \ref{sec:discussion}, the above estimate for $\delta q$ is expected to provide a standard uncertainty for $q$. 
An expanded uncertainty is obtained through multiplication of this value by a coverage factor.

\section{Application to real binary data} \label{sec:application}
We apply our proposed method to real binary data with spectroscopic mass ratios. 
We first surveyed the binaries in the catalog by \citet{Latkovic2021-ApJS} and identified 159 systems with entries of spectroscopic mass ratios. 
We then searched the TESS and Kepler archival data for their LCs, using the Python package Lightkurve \citep{Lightkurve2018-code}; 
LCs for 126 binaries were found. 
When two or more LCs were found for a binary, the better-quality one was selected in such a manner that its derivatives were less noisy and smoother than the other one. 
Apparent outliers in the LCs were removed in advance. 

We found that 64 of the 126 binaries showed statistically significant symmetric double peaks (see Section \ref{sec:key-value}) in the second derivatives of their LCs. 
Note that among the 64 samples, only two of them (KIC 10618253 and RZ Tau) were taken from the Kepler data and the rest were from the TESS data. 
In Latkovi{\'c}'s catalog, exactly half of the 64 binaries were of the W type, and half were of the A type.
These two categories are subtypes of the W Ursae Majoris (W UMa) binaries introduced by \citet{Binnendijk1970-VA}.

\begin{figure}
\centering
\includegraphics[width=0.48\textwidth]{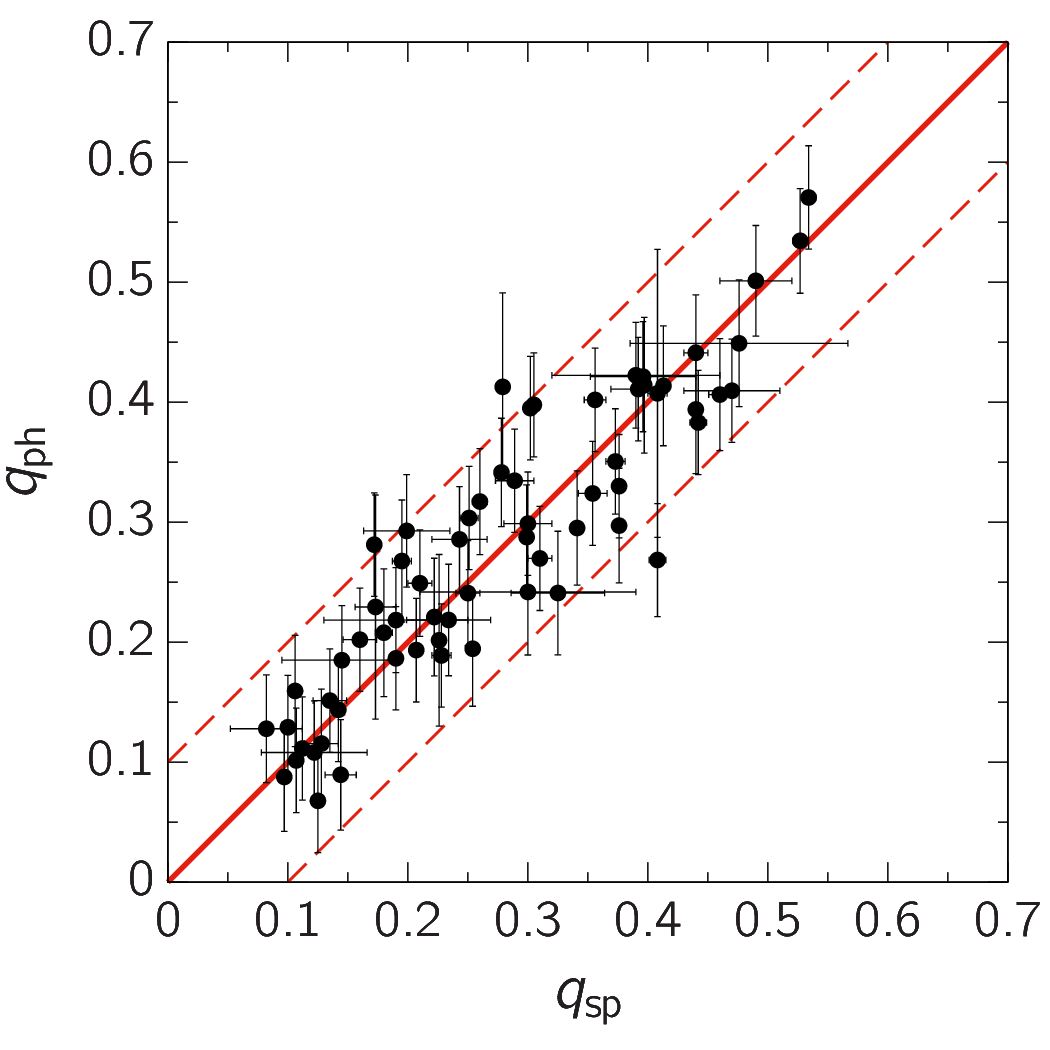}
\caption{Comparison between our photometric and the literature's spectroscopic mass ratios. 
The red solid and broken lines show $q_{\mathrm{ph}} = q_\mathrm{sp}$ and $q_{\mathrm{ph}} = q_\mathrm{sp} \pm 0.1$, respectively. 
\label{fig:qsp-qph}}
\end{figure}
\section{Results and Discussion} \label{sec:discussion}
\subsection{Comparison with spectroscopic mass ratios}
Figure \ref{fig:qsp-qph} shows a comparison between our mass ratio estimates ($q_\mathrm{ph}$) and corresponding spectroscopic mass ratios ($q_\mathrm{sp}$), 
the latter of which are compiled from the relevant papers listed in Latkovi{\'c}'s catalog. 
The standard deviation of $(q_\mathrm{ph}-q_\mathrm{sp})$ is 0.051, which is 0.008 larger than that for the synthesized LCs. 
This difference should originate from observational errors, which do not exist in the synthesized LCs. 
A slight difference in the standard deviation of $(q_\mathrm{ph}-q_\mathrm{sp})$ is found between the subtypes of W UMa binaries. 
The W- and A-type systems have standard deviations of 0.048 and 0.053, respectively. 
One possible factor contributing to this difference should be the variation in the uncertainties of spectroscopic mass ratios: 
specifically, the mean uncertainties for the W- and A-type systems were 0.015 and 0.017, respectively. 

The estimated mass ratios for 95\% of the binaries are within the range $q_\mathrm{sp} \pm 0.1$, which is $\sim $ four percentage points smaller than that for the synthesized LCs. 
This decrease should also be due to the observational errors mentioned above. 
Indeed, the estimated mass ratios for 67\% of the binaries are consistent with their spectroscopic values. 
This percentage is roughly the same as the probability (i.e., 68\%) that the true value falls within the confidence interval for the standard uncertainty. 
Consequently, the application to real binaries demonstrates that our method is practical for estimating the mass ratios of overcontact binaries and provides reasonable uncertainty values.

\subsection{Background}
\begin{figure}
\centering
\includegraphics[width=0.25\textwidth]{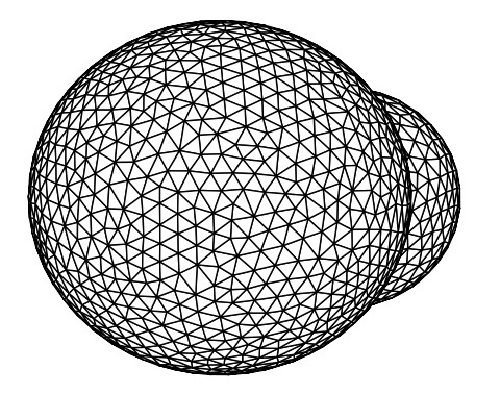}
\caption{
View of a binary from an observer at $t=t'_{32}=0.45$ day.
The parameters of the binary are the same as those shown in Figure \ref{fig:LC-diff}. 
\label{fig:binary}}
\end{figure}

We also discuss the background of why the proposed method works. 
It has been demonstrated that photometric mass ratios are accurately determined when overcontact binaries show total annular eclipses \citep[e.g.,][]{Pribulla2003-CoSka, Terrell2005-ApSS}. 
In our synthesized LC samples, condition (1) (Section \ref{sec:key-value}) effectively selects binaries with total annular eclipses. 
In addition, a photometric mass ratio is well determined if the ratio of the radii is known \citep{Terrell2005-ApSS}. 
Figure \ref{fig:binary} shows a schematic view of a binary from an observer at the key time $t'_{32}$. 
This is a moment immediately after the ratio of change in the luminosity becomes the minimum. 
In other words, it is a moment immediately after the common chord of the projected surfaces (i.e., circles) of the two stars becomes equal to the diameter of the smaller star, which is the maximum value that the common chord can take. 
Accordingly, the time interval $t_{32}-t'_{32}$ is a decreasing function of the relative size of the smaller star and, as such, embodies information on the ratio of the radii of the stars in a binary. 
Consequently, our method is expected to satisfy the conditions for accurately determining mass ratios and works well. 

\begin{figure*}
\centering
\includegraphics[width=0.95\textwidth]{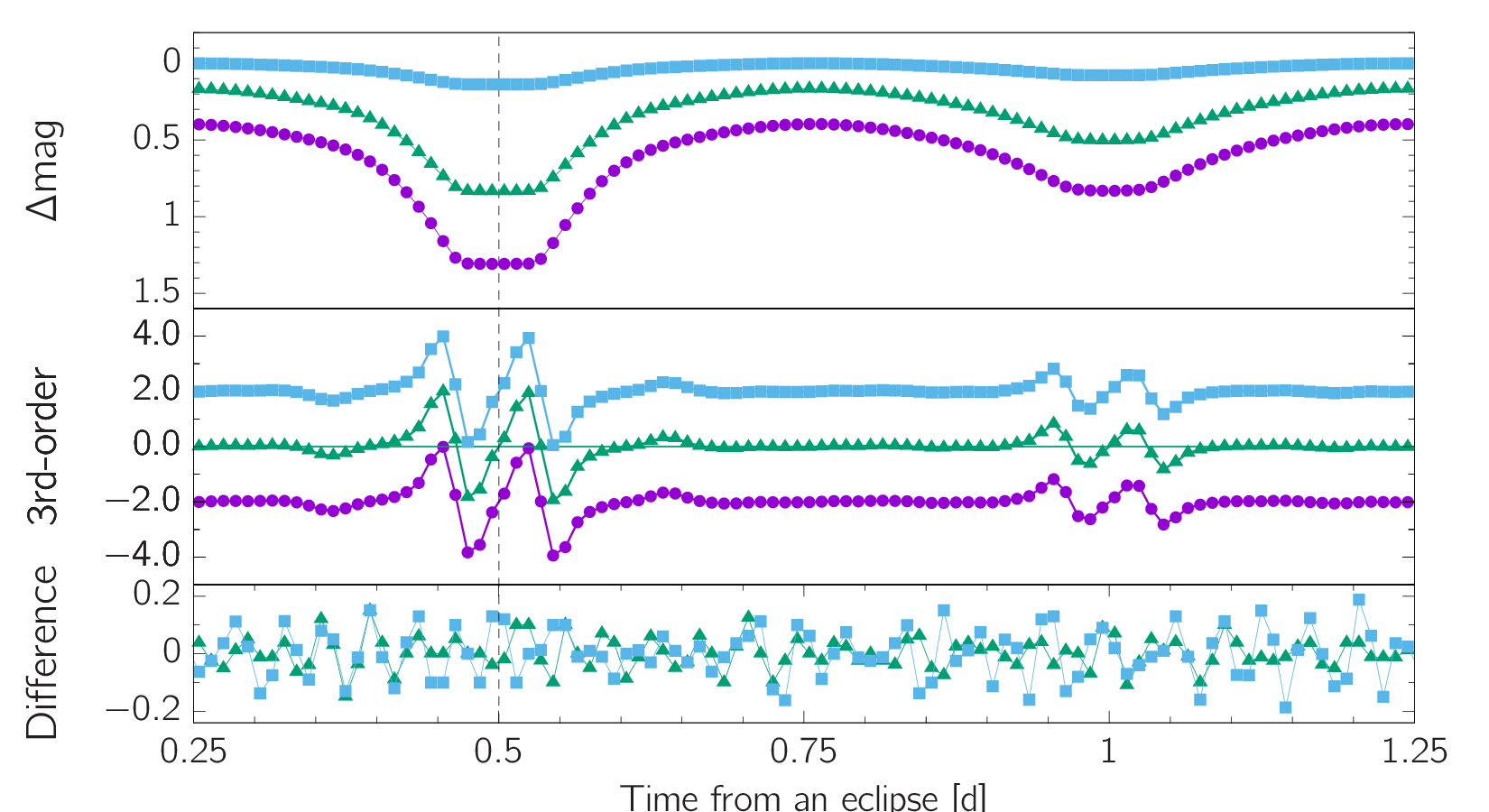}
\includegraphics[width=0.95\textwidth]{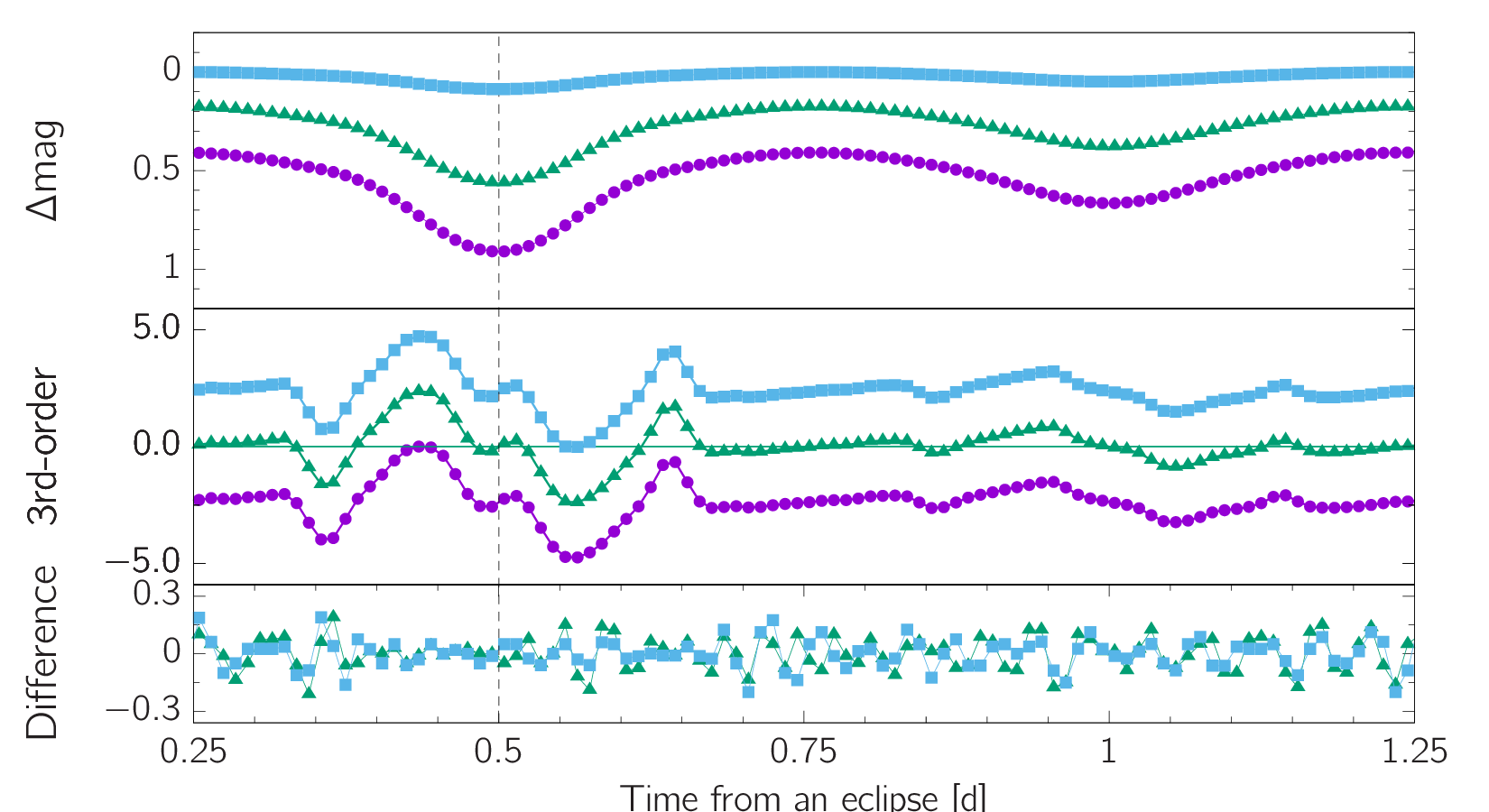}
\caption{Comparison of synthesized LCs (uppermost panel) and their third derivatives (middle panel) for overcontact binaries with $l_3 = 0$ (purple circles), $0.2$ (green triangles), and $0.8$ (blue squares), where $l_3$ is the ratio of the third light to the total light. 
The top figure corresponds to binaries with $i=90^\circ$ and $q=0.35$, while the bottom figure corresponds to binaries with $i=70^\circ$ and $q=0.15$. 
The other binary parameters are the same as those shown in Figure \ref{fig:LC-diff}. 
The LCs are represented in magnitudes, and the third derivatives are in units of $10^4$ and $10^3$ W m$^{-2}$ day$^{-3}$ for the binaries with $i=90^\circ$ and $70^\circ$, respectively. 
The top ($l_3=0.8$) and bottom ($l_3=0$) curves in the middle panels are shifted by $\pm 2.0\times 10^4$ and $\pm 2.0\times 10^3$ W m$^{-2}$ day$^{-3}$ for the binaries with $i=90^\circ$ and $70^\circ$, respectively. 
The lowermost panel shows the differences between the third derivatives of LCs with and without a third light, where the units are in W m$^{-2}$ day$^{-3}$. 
\label{fig:LC-comp}}
\end{figure*}
\subsection{Influence of third light}
The presence of light from the third body decreases the amplitudes of the variations in the LC of the eclipsing binary. 
In this section, we examine the influence of this third light on our mass ratio estimate. 

Figure \ref{fig:LC-comp} displays the LCs and their third derivatives for cases in which the ratio ($l_3$) of the third light to the total light (i.e., the light from the two component stars in the binary and from the third star) of the binary is 0, 0.2, and 0.8. 
The top (bottom) figure corresponds to binaries with $i=90^\circ$ and $q=0.35$ ($i=70^\circ$ and $q=0.15$).
The lowermost panel in each figure shows the differences between the third derivatives of LCs with and without a third light. 
Even though the amplitudes of the variations in the LCs differ between cases with and without third light, their third derivatives show good agreement. 
Furthermore, when using Equations (\ref{eq:q-reg}) and (\ref{eq:uncertainty}) to calculate the mass ratio and its uncertainty for all three LCs ($l_3=0$, $0.2$, and $0.8$), they all yield the values $0.310\pm 0.043$ and $0.188\pm 0.043$ for the binaries with $q=0.35$ and $0.15$, respectively; that is, all three values are in agreement.
Therefore, even when a constant third light contributes to the LC of an eclipsing binary, it does not affect the mass ratio estimate provided by our method.

\section{Conclusions} \label{sec:conclusion}
We have proposed a new method for estimating the mass ratio of an overcontact binary. 
The proposed method is based on our finding that the mass ratio correlates with the time interval between two local extrema in the third derivative of the LC. 
Our method requires only the calculation of the numerical derivatives of the LC and the measurement of the time interval. 
Unlike the widely used $q$-search method, our method does not require an iterative procedure. 
With this method, we can also derive a reasonable uncertainty value for the estimated mass ratio. 
An application to real binaries demonstrated that the estimated mass ratios for $\sim $67\% of the samples agreed with their spectroscopic mass ratios within the estimated uncertainties; 
the errors for $\sim $ 95\% of them are within $\pm 0.1$. 
In addition, this work also indicates that the derivative of the LC of an eclipsing binary might potentially embody important information on the properties of the binary. 

Our method is suitable for the LCs of overcontact eclipsing binaries with an adequate number of flux measurements and sufficient photometric precision. 
When a photometric mass ratio is desired to be obtained, such as in the absence of a radial velocity curve, our method can be useful. 
A mass ratio estimate should be helpful for the preliminary analyzing of an overcontact binary and for gaining insights into its properties. 

Recent surveys with high-precision time-series photometry, such as Kepler and TESS, have discovered numerous overcontact eclipsing binaries. 
Moreover, the number of samples is only expected to rise in the future. 
Currently, it is difficult to obtain radial velocity curves for most of the discovered eclipsing binaries without investing a significant, or even unrealistic, amount of resources. 
By contrast, our method is handy for estimating their mass ratios with reasonably trustworthy uncertainties for most of the discovered binaries using the photometric data already available. 
It should also allow us to advance statistical studies of overcontact binaries. 

\vspace{5mm}
The author would like to thank the anonymous referee for comments and suggestions, which have been helpful in improving the paper. 
This paper includes data collected by the TESS mission. 
Funding for the TESS mission is provided by the NASA's Science Mission Directorate. 
This paper includes data collected by the Kepler mission and obtained from the MAST data archive at the Space Telescope Science Institute (STScI). 
Funding for the Kepler mission is provided by the NASA Science Mission Directorate. 
STScI is operated by the Association of Universities for Research in Astronomy, Inc., under NASA contract NAS 5–26555.

\bibliographystyle{aasjournal}

\begin{thebibliography}{}
\expandafter\ifx\csname natexlab\endcsname\relax\def\natexlab#1{#1}\fi
\providecommand{\url}[1]{\href{#1}{#1}}
\providecommand{\dodoi}[1]{doi:~\href{http://doi.org/#1}{\nolinkurl{#1}}}
\providecommand{\doeprint}[1]{\href{http://ascl.net/#1}{\nolinkurl{http://ascl.net/#1}}}
\providecommand{\doarXiv}[1]{\href{https://arxiv.org/abs/#1}{\nolinkurl{https://arxiv.org/abs/#1}}}

\bibitem[{{Binnendijk}(1970)}]{Binnendijk1970-VA}
{Binnendijk}, L. 1970, Vistas in Astronomy, 12, 217,
  \dodoi{10.1016/0083-6656(70)90041-3}

\bibitem[{{Conroy} {et~al.}(2020){Conroy}, {Kochoska}, {Hey}, {Pablo},
  {Hambleton}, {Jones}, {Giammarco}, {Abdul-Masih}, \&
  {Pr{\v{s}}a}}]{Conroy2020-ApJS}
{Conroy}, K.~E., {Kochoska}, A., {Hey}, D., {et~al.} 2020, \apjs, 250, 34,
  \dodoi{10.3847/1538-4365/abb4e2}

\bibitem[{{Kopal}(1959)}]{Kopal1959-cbs}
{Kopal}, Z. 1959, {Close binary systems}

\bibitem[{{Latkovi{\'c}} {et~al.}(2021){Latkovi{\'c}}, {{\v{C}}eki}, \&
  {Lazarevi{\'c}}}]{Latkovic2021-ApJS}
{Latkovi{\'c}}, O., {{\v{C}}eki}, A., \& {Lazarevi{\'c}}, S. 2021, \apjs, 254,
  10, \dodoi{10.3847/1538-4365/abeb23}

\bibitem[{{Lightkurve Collaboration} {et~al.}(2018){Lightkurve Collaboration},
  {Cardoso}, {Hedges}, {Gully-Santiago}, {Saunders}, {Cody}, {Barclay}, {Hall},
  {Sagear}, {Turtelboom}, {Zhang}, {Tzanidakis}, {Mighell}, {Coughlin}, {Bell},
  {Berta-Thompson}, {Williams}, {Dotson}, \& {Barentsen}}]{Lightkurve2018-code}
{Lightkurve Collaboration}, {Cardoso}, J.~V.~d.~M., {Hedges}, C., {et~al.}
  2018, {Lightkurve: Kepler and TESS time series analysis in Python},
  Astrophysics Source Code Library.
\newblock \doeprint{1812.013}

\bibitem[{{Lucy}(1968)}]{Lucy1968-ApJ877}
{Lucy}, L.~B. 1968, \apj, 153, 877, \dodoi{10.1086/149712}

\bibitem[{{Mazeh} \& {Goldberg}(1992)}]{Mazeh1992-ApJ}
{Mazeh}, T., \& {Goldberg}, D. 1992, \apj, 394, 592, \dodoi{10.1086/171611}

\bibitem[{{Osaki}(1965)}]{Osaki1965-PASJ}
{Osaki}, Y. 1965, \pasj, 17, 97

\bibitem[{{Pribulla} {et~al.}(2003){Pribulla}, {Kreiner}, \&
  {Tremko}}]{Pribulla2003-CoSka}
{Pribulla}, T., {Kreiner}, J.~M., \& {Tremko}, J. 2003, Contributions of the
  Astronomical Observatory Skalnate Pleso, 33, 38

\bibitem[{{Rucinski}(2001)}]{Rucinski2001-AJ1007}
{Rucinski}, S.~M. 2001, \aj, 122, 1007, \dodoi{10.1086/321153}

\bibitem[{{Terrell}(2022)}]{Terrell2022-Galax}
{Terrell}, D. 2022, Galaxies, 10, 8, \dodoi{10.3390/galaxies10010008}

\bibitem[{{Terrell} \& {Wilson}(2005)}]{Terrell2005-ApSS}
{Terrell}, D., \& {Wilson}, R.~E. 2005, \apss, 296, 221,
  \dodoi{10.1007/s10509-005-4449-4}

\bibitem[{{Trimble}(1990)}]{Trimble1990-MNRAS}
{Trimble}, V. 1990, \mnras, 242, 79, \dodoi{10.1093/mnras/242.1.79}

\bibitem[{{Wilson}(1994)}]{Wilson1994-PASP}
{Wilson}, R.~E. 1994, \pasp, 106, 921, \dodoi{10.1086/133464}

\end{thebibliography}

\end{document}